\documentclass[12pt, a4paper]{article}
\usepackage[margin=1in]{geometry}
\usepackage{amsfonts, amsmath, amssymb, bm}
\usepackage{mathtools}
\usepackage{tabu}
\usepackage[utf8x]{inputenc}
\usepackage{enumerate}
\usepackage{enumitem}
\usepackage{tgcursor}
\usepackage{xcolor}
\usepackage{graphicx}
\usepackage{placeins}
\usepackage{listings}
\usepackage{stackengine}
\usepackage{caption}
\usepackage{subcaption}
\usepackage{mathptmx}
\usepackage{setspace}
\usepackage{footnote}
\usepackage{comment}
\usepackage{footmisc}
\usepackage{hyperref}
\usepackage{authblk}
\usepackage{eucal}
\usepackage{verbatim}
\usepackage{mathrsfs}
\usepackage{natbib}
\usepackage[colorinlistoftodos]{todonotes}
\usepackage{threeparttable}
\usepackage{algorithm2e}
\SetKwProg{init}{Initialize:}{}{}
\RestyleAlgo{ruled}

\renewcommand{\vec}[1]{\boldsymbol{#1}}

\newcommand{\vA}{\mathbf{A}}
\newcommand{\vb}{\mathbf{b}}
\newcommand{\vB}{\mathbf{B}}

\newcommand{\vf}{\mathbf{f}}
\newcommand{\vF}{\mathbf{F}}
\newcommand{\vg}{\mathbf{g}}

\newcommand{\vH}{\mathbf{H}}

\newcommand{\vI}{\mathbf{I}}

\newcommand{\vm}{\mathbf{m}}
\newcommand{\vM}{\mathbf{M}}

\newcommand{\vr}{\mathbf{r}}

\newcommand{\vs}{\mathbf{s}}

\newcommand{\vu}{\mathbf{u}}
\newcommand{\vU}{\mathbf{U}}
\newcommand{\vv}{\mathbf{v}}
\newcommand{\vV}{\mathbf{V}}

\newcommand{\vW}{\mathbf{W}}

\newcommand{\vepsilon}{\boldsymbol{\epsilon}}

\newcommand{\veta}{\boldsymbol{\eta}}

\newcommand{\vgamma}{\boldsymbol{\gamma}}

\newcommand{\vomega}{\boldsymbol{\omega}}
\newcommand{\vOmega}{\boldsymbol{\Omega}}
\newcommand{\vphi}{\boldsymbol{\phi}}
\newcommand{\vPhi}{\boldsymbol{\Phi}}

\newcommand{\vpsi}{\boldsymbol{\psi}}
\newcommand{\vPsi}{\boldsymbol{\Psi}}

\newcommand{\vsigma}{\boldsymbol{\sigma}}
\newcommand{\vSigma}{\boldsymbol{\Sigma}}

\newcommand{\vtheta}{\boldsymbol{\theta}}
\newcommand{\vTheta}{\boldsymbol{\Theta}}

\hypersetup{
    colorlinks=true,
    linkcolor=blue,
    urlcolor=blue,
    citecolor=black
}

\title{\vspace{-2cm}\LARGE\centering\normalfont{A Review of Data-Driven Discovery for Dynamic Systems}}
\date{\vspace{-10mm}}

\author[1,*]{Joshua S. North}
\author[2]{Christopher K. Wikle}
\author[3]{Erin M. Schliep}
\affil[1]{Earth and Environmental Sciences, Lawrence Berkeley National Laboratory, Berkeley, CA, 1 Cyclotron Road}
\affil[2]{Department of Statistics, University of Missouri, Columbia, MO, 146 Middlebush Hall}
\affil[3]{Department of Statistics, North Carolina State University, Raleigh, NC, 2311 Stinson Drive}
\affil[*]{Corresponding author: jsnorth@lbl.gov}

\begin{document}

\maketitle

\begin{abstract}

Many real-world scientific processes are governed by complex nonlinear dynamic systems that can be represented by differential equations.
Recently, there has been increased interest in learning, or discovering, the forms of the equations driving these complex nonlinear dynamic system using data-driven approaches.
In this paper we review the current literature on data-driven discovery for dynamic systems.
We provide a categorization to the different approaches for data-driven discovery and a unified mathematical framework to show the relationship between the approaches.
Importantly, we discuss the role of statistics in the data-driven discovery field, describe a possible approach by which the problem can be cast in a statistical framework, and provide avenues for future work.

    \begin{center}
    \end{center}
    
\end{abstract}

\section{Introduction}\label{sec:intro}

Recently there has been a push from within the computer science, physics, applied mathematics, and statistics to learn the governing equations in complex dynamic systems parameterized through dynamic equations (DE).
There are a variety of reasons researchers may want to know the underlying laws driving a system -- to reinforce their assumptions, uncover extra information about the system, or to produce a more realistic mathematical representation for the system.
Historically, scientists have relied on their ability to represent physical systems using mathematical equations in the form of DEs.
Dating back to at least the inference of equations describing the motion of orbital bodies around the sun based on the positions of celestial bodies \citep{Legendre1806,Gauss1809}, DEs have been used to model the evolution of complex processes (e.g., the use of susceptible, infected, recovered models for epidemics), and have become ubiquitous across virtually every area of science and engineering.
Here, we review some of the methods used to discover the governing equations driving complex, potentially nonlinear, processes, often referred to as \textit{data-driven discovery}.

Consider the general DE describing the evolution of a continuous process $\{\vu(\vs, t): \vs \in D_s, t \in D_t\},$ 
\begin{align}\label{eqn:pde}
    \vu_{t^{(J)}}(\vs, t) = 
    M\left(\vu(\vs, t), \vu_x(\vs, t), \vu_y(\vs, t), ..., \vu_{t^{(1)}}(\vs, t), ..., \vu_{t^{(J-1)}}(\vs, t), \vomega(\vs, t)
    \right),
\end{align}
where the vector $\vu(\vs, t) \in \mathbb{R}^N$ denotes the state of the system at location $\vs$ and time $t$, $\vu_{t^{(j)}}(\vs, t)$ is the $j$th order temporal derivative of $\vu(\vs, t)$, $J$ denotes the highest order of the temporal derivative, $M(\cdot)$ represents the (potentially nonlinear) evolution function, and $\vomega(\vs, t)$ represents any covariates that might be included in the system.
We will denote partial derivatives by a subscript; that is $\frac{\partial \vu}{\partial x} = \vu_x$ and $\frac{\partial \vu}{\partial t} = \vu_t$, for example.
Here, $N$ is the number of components in the system (e.g., $\vu(\vs,t) = [u(\vs, t, 1), u(\vs, t, 2), ..., u(\vs, t, N)]'$, sometimes called the system state), $\vs \in \{\vs_1, ..., \vs_S\} = D_s$ is a discrete location in the domain with $|D_s| = S$, and $t \in \{1, ..., T\} = D_t$ is the realization of the system at discrete times where $|D_t| = T$.
Equation (\ref{eqn:pde}) is composed of partial derivatives of the system with $D_s \in \mathbb{R}^2$ and $\vs = (x,y)$ (although this can be simplified to $\mathbb{R}^1$ with $\vs = x$ or generalized to higher dimensions) and is often referred to as a partial differential equation (PDE).
Removing the spatial component from (\ref{eqn:pde}) results in a temporal ordinary differential equation (ODE),
\begin{align}\label{eqn:ode}
    \vu_{t^{(J)}}(t) = 
    M\left(\vu(t), \vu_{t^{(1)}}(t), ..., \vu_{t^{(J-1)}}(t), \vomega(t)
    \right),
\end{align}
where $M$ is composed solely of derivatives of the components in time (i.e., no partial derivatives). 
This review will focus on methods to discover the evolution function $M$ for both PDEs (\ref{eqn:pde}) and ODEs (\ref{eqn:ode}).

The goal of data-driven discovery is to learn the governing equation(s) in (\ref{eqn:pde}) and (\ref{eqn:ode}) -- specifically the (non)linear function $M$ -- having only observed noisy realizations of the true process $\vu$ (i.e., true derivatives are unknown).
Broadly, we group the approaches used for data-driven discovery into three categories
-- classical sparse methods, classical symbolic methods, and deep modeling methods using either symbolic or sparse regression techniques -- 
but recognize other categorization is possible.
The first approach uses sparse regression where a library of potential solutions are proposed and the correct solution set is obtained by regularization based techniques, resulting in a sparse solution.
The second uses symbolic regression where the solution is learned, or generated, through the estimation procedure.
The third uses deep models to facilitate the discovery process of the previous two approaches (e.g., symbolic regression using deep models).
As this is an active area of research, we refer the reader to the special issue \citet[][]{PTRSA} for emerging areas of research and applications.


While less common than the deterministic counterparts, methods to quantify uncertainty in the discovered equations have been proposed.
However, these methods generally do not account for uncertainty in the observed data, missing a vital piece of the statistical puzzle.
We draw parallels between traditional statistical models and data-driven discovery, discussing how statistical models can be formulated for data-driven discovery and highlighting possible improvments to the methods.

The remainder of the paper is organized as follows.
In Section \ref{sec:sparse} we review sparse regression methods for data-driven discovery, which are sub-categorized into deterministic and probabilistic approaches.
In Section \ref{sec:symbolic} we review symbolic methods for data-driven discovery.
In Section \ref{sec:deep} we review deep modeling approaches for data-driven discovery, which are sub-divided into methods approximating and discovering the underlying dynamics.
In Section \ref{sec:psm} we show how the problem can be formulated in a statistical paradigm and in Section \ref{sec:pdetection} we review a possible method of data-driven discovery using a fully probablistic approach.
Section \ref{sec:discussion} concludes the paper.

\section{Sparse Regression}\label{sec:sparse}

Sparse regression approaches for dynamic discovery of ODEs and PDEs are fundamentally the same.
We formulate the general approach using (\ref{eqn:pde}), noting that the approach for (\ref{eqn:ode}) is equivalent but with only one spatial location (i.e., $S = 1$).
First, consider rewriting (\ref{eqn:pde}) as a linear (in parameters) system
\begin{align*}
    \vu_{t^{(J)}}(\vs, t) = \vf\left(\vu(\vs, t), ...\right) \vM,
\end{align*}
where $\vM$ is a $D \times N$ \textit{sparse} matrix of coefficients and $\vf(\cdot)$
is a vector-valued nonlinear transformation function of length $D$ termed the \textit{feature library}.
The input of the arguments for $\vf(\cdot)$ are general and contain terms that \textit{potentially} relates to the system (e.g., advection term, polynomial terms, interactions).
Sparse identification seeks to identify relevant terms of $\vM$, thereby identifying the components of $\vf$ that drive the system and discovering the governing dynamics.

Denote the matrix of all data (all components at all time points) for the $j$th derivative of the system as
\begin{align*}
    \vU_{t^{(j)}} & = 
    \begin{bmatrix}
        u_{t^{(j)}}(\vs_1, 1, 1) & u_{t^{(j)}}(\vs_1, 1, 2) & \cdots & u_{t^{(j)}}(\vs_1, 1, N) \\
        u_{t^{(j)}}(\vs_1, 2, 1) & u_{t^{(j)}}(\vs_1, 2, 2) & \cdots & u_{t^{(j)}}(\vs_1, 2, N) \\
        \vdots              & \vdots              & \vdots                       \\
        u_{t^{(j)}}(\vs_S, T, 1) & u_{t^{(j)}}(\vs_S, T, 2) & \cdots & u_{t^{(j)}}(\vs_S, T, N) \\
    \end{bmatrix}.
\end{align*}
The response matrix is $\vU_{t^{(J)}}$ of size $(S T) \times N$ and we generically denote the feature library as
\begin{align*}
    \vF = \left[ \vec{1}, \vU_{t^{(0)}}, ..., \vU_{t^{(J)}}, \vU_x, \vU_y, \vU_{xx}, ... , \vOmega \right].
\end{align*}
where $\vOmega$ are the associated covariates indexed in space and time and $\vF$ is a $ (ST) \times D$ matrix.
The library may also contain interactions of the components, partial derivatives, and covariates.
We can write the linear system
\begin{align}\label{eqn:ode_linear_system}
    \vU_{t^{(J)}} = \vF \vM,
\end{align}
whereby identifying the terms of $\vM$ that are non-zero, the DE is identified.

The derivatives of the system are rarely observed (i.e., only $\vU_{t^{(0)}}(t)$ is measured).
To obtain derivatives in space and time, numerical techniques are used to approximate the derivatives.
There are multiple methods to approximate derivatives numerically, and the choice of approximation has the potential to impact the discovered equation  \citep{DeSilva2020}.
Originally, a finite difference approach was suggested, but this approach is sensitive to noise \citep{Chartrand2011}.
When measurement noise is present, data are either smoothed \textit{a priori} and then derivatives are computed, or derivatives are computed using either total variation regularization \citep{Chartrand2011} or polynomial interpolation \citep{Knowles2012}.

Due to both the numerical approximation of the derivative and the potential for noise in the observed data, (\ref{eqn:ode_linear_system}) does not hold exactly.
Instead,
\begin{align}\label{eqn:sindy_system}
    \vU_{t^{(J)}} = \vF \vM + \vepsilon,
\end{align}
where $\vepsilon \overset{i.i.d.}{\sim} N(\vec{0}, \sigma^2\vI_N)$ and $\sigma^2$ is the variance associated with the model approximation and the numerical differentiation.
To induce sparsity, and thereby identify the relevant terms governing the system, solutions to (\ref{eqn:sindy_system}) of the form
\begin{align}\label{eqn:penalized_likelihood}
    \vM = \underset{\widehat{\vM}}{argmin}\|\vU_{t^{(J)}} - \vF \widehat{\vM}\|^2_2 + Pen_{\theta}(\widehat{\vM}),
\end{align}
are sought, where $Pen_{\theta}(\widehat{\vM})$ generically denotes some penalty term based on parameters $\theta$ (i.e., $Pen_{\theta}(\widehat{\vM}) = \lambda \|\widehat{\vM}\|_1$ where $\theta = \lambda$ for the LASSO penalty).

\subsection{Deterministic Approaches}\label{sec:sparse_deterministic}

The majority of deterministic approaches are composed of three steps -- denoising and differentiation, construction of a feature library, and sparse regression.
Assuming data have been properly differentiated and a library has been proposed, the deterministic approach seeks solutions of the form (\ref{eqn:penalized_likelihood}).
The original sparse regression approach to data-driven discovery, \textit{Sparse Identification of Nonlinear Dynamics} \citep[SINDy;][]{Brunton2016}, uses sequential threshold least-squares (STLS; Algorithm \ref{alg:SINDy}) to discover the governing terms for ODEs.
While the original paper does not discuss the algorithm in terms of a penalty term, STLS has been shown to be equivalent to the $\ell_0$ penalty, $Pen_{\theta}(\widehat{\vM}) = \|\widehat{\vM}\|_0$ \citep{Zhang2019}, which removes values of $\vM$ less than some pre-specified threshold $\kappa$.
That is, at each iteration of the minimization procedure, values of $\vM < \kappa$ are set to zero and the remaining values of $\vM$ are re-estimated.
In the original implementation, the algorithm was only iterated 10 times, but a stopping criteria (e.g., change in loss or identified parameters) could be used.
In this manner, a sparse solution set is obtained.

In the literature, SINDy is illustrated on a variety of simulated ODE problems with varying amounts of noise.
The examples used generally contain a lot of observations (on the order of hundreds of thousands), and it is unclear the impact of noise if a smaller number of observations is considered.
In contrast to the symbolic approaches discussed in Section \ref{sec:symbolic}, SINDy can be fit rather quickly.
However, a drawback of the approach is the sensitivity to the thresholding parameter and the dependence on the method approximating the derivative.

To extend SINDy to PDEs, \citet{Rudy2017} propose Sequential Threshold Ridge Regression (STRidge, Algorithm \ref{alg:pde_find}), a variant to STLS.
Due to the correlation present in $\vF$ for data pertaining to PDEs, STLS is insufficient at finding a sparse solution set.
Instead, STRidge uses the same iterative technique as STLS, where values of $\vM < \kappa$ are set to zero at each iteration, but with the addition of the penalty term $Pen_{\theta}(\widehat{\vM}) = \lambda \|\widehat{\vM}\|_2^2$.
Cross-validation is then used to find the optimal values for $\kappa$ and $\lambda$.
The effectiveness of STRidge is shown on multiple simulated data sets with varying noise.
Again, in comparison to the symbolic counterparts, the algorithm is quick, but still dependent on the method used to approximate the derivative.

STRidge can be adapted to allow for parametric PDEs by grouping terms either spatially or temporally \citep{Rudy2019}.
To incorporate parametric PDEs in \ref{eqn:ode_linear_system}, the coefficients now vary in space or time (i.e., $\vM(\vs)$ or $\vM(t)$) and $\vF$ is constructed as a block diagonal matrix of the appropriate form (e.g., either in space or time).
Similar to the group LASSO \citep{Meier2008}, coefficients are assigned to a collection $g \in \mathcal{G}$ by grouping the same location in space over the entire time domain (e.g., $g \equiv \vs$ and $\mathcal{G} \equiv D_S$) or the same time point over the whole spatial domain (e.g., $g \equiv t$ and $\mathcal{G} \equiv D_T$).
Within the STRidge algorithm all coefficients with the same group index are set to zero if $\|\vM(g)\|_2 < \kappa$.
In this manner, the same dynamics are identified across space and time and only the coefficient estimate is allowed to vary in space or time.

\citet{Champion2020} propose a robust unifying algorithm (Algorithm \ref{alg:sr3}) for the SINDy framework based on sparse relaxed regularized regression \citep[SR3;][]{Zhang2018b}.
SR3 introduces an auxiliary variable matrix $\vW$ within the penalization term, resulting in $Pen_{\theta}(\widehat{\vM}) = \lambda R(\vW) + \frac{1}{2\nu}\|\widehat{\vM} - \vW\|$, where $R(\cdot)$ is another penalization term (e.g., $\ell_1$).
The addition of the auxiliary variables provides a more favorable geometric surface to optimize \citep{Zhang2018b}.
SR3 is shown to be able to handle outliers (a potential issue when numerically differentiating noisy data), accommodate parametric formulations, and allow for physical constraints in the library.

While not discussed in detail here, there are other applications and approaches of DE discovery using sparse methods within the literature.
Applying SINDy to stochastic differential equations \citep{Boninsegna2018} and systems where the dynamics evolve on a different coordinate system \citep{Champion2019} further increase the SINDy applicability.
Instead of using finite differences or total variation regularization, \citet{Schaeffer2017} use spectral methods to compute spatial derivatives and the Douglas-Rachford algorithm \citep{Combettes2011} to find a sparse solution.
Further consideration of highly corrupt signals \citep{Tran2017}, convergence properties of the SINDy algorithm \citep{Zhang2019}, and the choice of denoising and differentiation methods \citep{Lagergren2020} have also received treatment within the literature.
For ease of use, SINDy and some related variants are available in the python package \textit{PySINDy} \citep{DeSilva2020}.

\subsection{Addressing Uncertainty}\label{sec:sparse_uncertainty}

Bayesian and bootstrapping approaches have been proposed to quantify uncertainty in the parameters for the sparse regression formulation of data-driven discovery.
These approaches seek to quantify the variability in the discovered equation and parameters of (\ref{eqn:sindy_system}).

\subsubsection{Bayesian Approach}\label{sec:bayes_uncertainty}

A penalized likelihood estimator of the form (\ref{eqn:penalized_likelihood}) can analogously be cast as the posterior mode in a Bayesian framework under the prior $p(\vM|\theta)$ where $Pen(\widehat{\vM})_{\theta} = \log p(\vM|\theta)$.
That is, (\ref{eqn:sindy_system}) can be formulated in the Bayesian framework where priors are put on $\vM$ and $\sigma^2$.
Instead of an optimization procedure, the Bayesian approach aims to sample from the joint posterior distribution
\begin{align*}
    p(\vM, \vsigma^2|\vF, \vU_{t^{(j)}}) \propto p(\vU_{t^{(j)}}|\vF, \vM, \vsigma^2)p(\vM|\theta)p(\vsigma^2|\theta)p(\theta),
\end{align*}
where $p(\vU_{t^{(j)}}|\vF, \vM, \vsigma^2)$ is the data likelihood (\ref{eqn:sindy_system}), and $p(\vM|\theta)$ and $p(\vsigma^2|\theta)$ are prior distributions for $\vM$ and $\vsigma^2$, respectively, and $p(\theta)$ is the prior distribution for the hyperparameters (or penalization parameters in (\ref{eqn:penalized_likelihood})).
To enforce a sparse solution set in a Bayesian framework, a regularization prior is placed on the parameter of interest, in this case $\vM$.
Further discussion comparing the sparse regression approach to a Bayesian formulation of the problem is explored by \citet{Niven2020}.

Using the Bayesian framework in an algorithmic setting, \citet{Zhang2018b} propose using the priors $p(m_d|\alpha_d) = \prod_{d=1}^D N(0, \alpha_d^{-1})$, $p(\sigma^2) = IG(a_s,b_s)$, and $p(\alpha_d^{-1}) = IG(a_a, b_a)$, where $N$ is the normal (Gaussian) distribution and $IG$ is the inverse Gamma distribution.
They estimate the parameters using a threshold sparse Bayesian regression algorithm, which maximizes the marginal likelihood instead of sampling from the full conditional distributions as in Markov chain Monte Carlo.
Their algorithm uses a hard thresholding parameter, similar to the deterministic sparse regression approaches, where at each iteration, values of the posterior $\vM(i,j) < \kappa$ are set to zero.
Their procedure assigns what they term ``error bars'' to their parameter estimates based on the ratio of the estimate for the posterior variance to the estimate for the posterior mean squared.
\citet{Zhang2018b} consider many of the same simulated ODEs and PDEs used to illustrate the deterministic approaches and provide error bars to the parameter estimates for these systems with varying amounts of measurement noise.
As an interesting application, \citet{Zanna2020} use this framework to discovery unknown equations in ocean mesoscale closures.

\citet{Hirsh2021} explore the use of two common Bayesian selection priors on system discovery and uncertainty quantification -- the continuous spike and slab \citep[i.e., stochastic search variable selection (SSVS);][]{Mitchell1988, George1993, George1997}, and the regularized horseshoe \citep{Carvalho2010, Piironen2017} -- calling the approach \textit{uncertainty quantification SINDy} (UQ-SINDy).
Their choice of priors are distinct in that SSVS is a mixture of two continuous mean zero Gaussian distributions and the horseshoe is part of the global-local shrinkage prior family.
For the SSVS prior, variables that are not to be included in the model are sampled from a mean zero Gaussian distribution with a small variance, rendering their effect on inference negligible, and variables that are to be included are sampled from a mean zero Gaussian distribution with a larger variance.
The posterior inclusion probability for a variable is the number of times it was sampled from the Gaussian with a large variance over the total number of samples.
In contrast, the horseshoe prior has a hyper-prior performing global shrinkage on all variables in conjunction with individual hyper-priors performing individual shrinkage.
To determine the probability a variable is included under the regularized horseshoe, the ratio of each element of the estimate of $\vM$ with no prior and with the horseshoe prior is computed, providing \textit{pseudo-inclusion probabilities} (i.e., not necessarily bounded by 0 and 1).
Using both of these priors, \citet{Hirsh2021} provide inclusion probabilities for multiple simulated ODE systems with varying amounts of noise and to the classic hare-lynx population data set \citep{Elton1942}.

However, these two approaches are limited in that the uncertainty being quantified is the uncertainty in the numerical approximation of the system (i.e., the numerical differentiation and de-noising).
That is, because the approximated derivative is, in fact, a single realization of the true derivative (which is unknown), the uncertainty estimates recovered by this approach are biased toward this single approximation of the derivative.
A more complete treatment of the problem would be to consider the derivative as a random process and account for uncertainty in the random process.

\citet{Yang2020} propose the use of Bayesian differentiable programming as a method by which to discover the dynamics and account for measurement uncertainty when estimating parameters.
Generally speaking, Bayesian differentiable programming uses a numerical solver (e.g., Runge-Kutta) to predict the state at a new time, and the loss between the predicted data and observed data is used to estimate parameters.
More precisely, letting $M_{\theta}(\vu(t))$ be the output of a numerical solver at time $t$, Bayesian differentiable programming aims to minimize $\sum \| \vu(t+\Delta t) - M_{\theta}(\vu(t))\|^2$, where $\Delta t$ does not need to be uniformly spaced.
The parameters are estimated using Hamiltonian Monte Carlo and differentiable programming is used to compute gradients within the Hamiltonian Monte Carlo algorithm.
By directly relating the observed data to the dynamics, measurement uncertainty is accounted for in the estimation procedure, providing a more thorough statistical treatment to the data-driven discovery problem.
The approach is illustrated on multiple simulated ODE systems with varying amounts of measurement noise.

\citet{Bhouri2022} extend the idea of Bayesian differential programming for data-driven discovery by using Gaussian process priors on the state variables to model temporal correlations and use NuralODEs \citep{Chen2019b} to perform numerical integration.
Additionally, they use the ``Finnish Horseshoe prior'' to impose variable shrinkage on the learned library, similar to \citet{Hirsh2021}.
The model parameters are estimated using Hamiltonian Monte Carlo and the No-U-Turn-Sampler \citep[NUTS;][]{Hoffman2014} algorithm to automatically calibrate model parameters.
Similar to \citet{Yang2020}, the approach is illustrated on multiple simulated ODE systems in addition to the human motion capture data.

\subsubsection{Bootstrap Approach}\label{sec:ensemble_uncertainty}

\citet{Fasel2021} propose two methods of bootstrapping (\ref{eqn:ode_linear_system}) -- either sampling rows of the data (i.e., space-time sampling) or sampling library terms (i.e., columns of $\vF$).
The first approach samples rows of the data with replacement and uses STRidge to estimate the parameters in the model $q$ times.
In the second approach, the columns of $\vF$ are sampled \textit{without} replacement to create $q$ data sets, and again STRidge is used to estimate parameters.
For both methods, the $q$ models are then averaged and coefficients with an inclusion probability below some pre-specified value are set to zero.
Uncertainty is quantified by the inclusion probability and the distribution of values obtained from the $q$ different estimates.
However, as with \citet{Hirsh2021}, the uncertainty associated with the observed data is not considered and the numerical approximation to the derivative is treated as an observation of the derivative, limiting uncertainty quantification.
The method is illustrated on multiple simulated ODE and PDE systems with varying noise and applied to the classic hare-lynx population data set.

\section{Symbolic Regression}\label{sec:symbolic}

Symbolic regression is a type of regression that searches over mathematical expressions (e.g., $+, -, \times$) to find the optimal model for a given data set \citep{Wang2019}.
This approach differs from classical regression where the model structure is fixed and a set of parameters are estimated.
One of the main challenges underlying symbolic regression is that there are an infinite number of combinations of expressions that can be used to fit any particular data set.
\textit{Genetic programming} is used to efficiently search over the possible model structures \citep{Willis1997, Koza1993, Koza1997} and regression techniques are used to determine coefficient values given the model structure.
Genetic programming follows Darwin's theory of evolution, selecting the ``fittest'' solution that is the product of generations of evolution (i.e., iterating through an algorithm).
Here, we give a brief overview of genetic programming and its roll in symbolic regression and subsequently data-driven discovery of dynamics.
For a more detailed overview of genetic programming, see \citet[Chapter~4,][]{Minnebo2011} and \citet{Garg2012}.

\begin{figure}
    \centering
    \includegraphics[width = 0.8\linewidth]{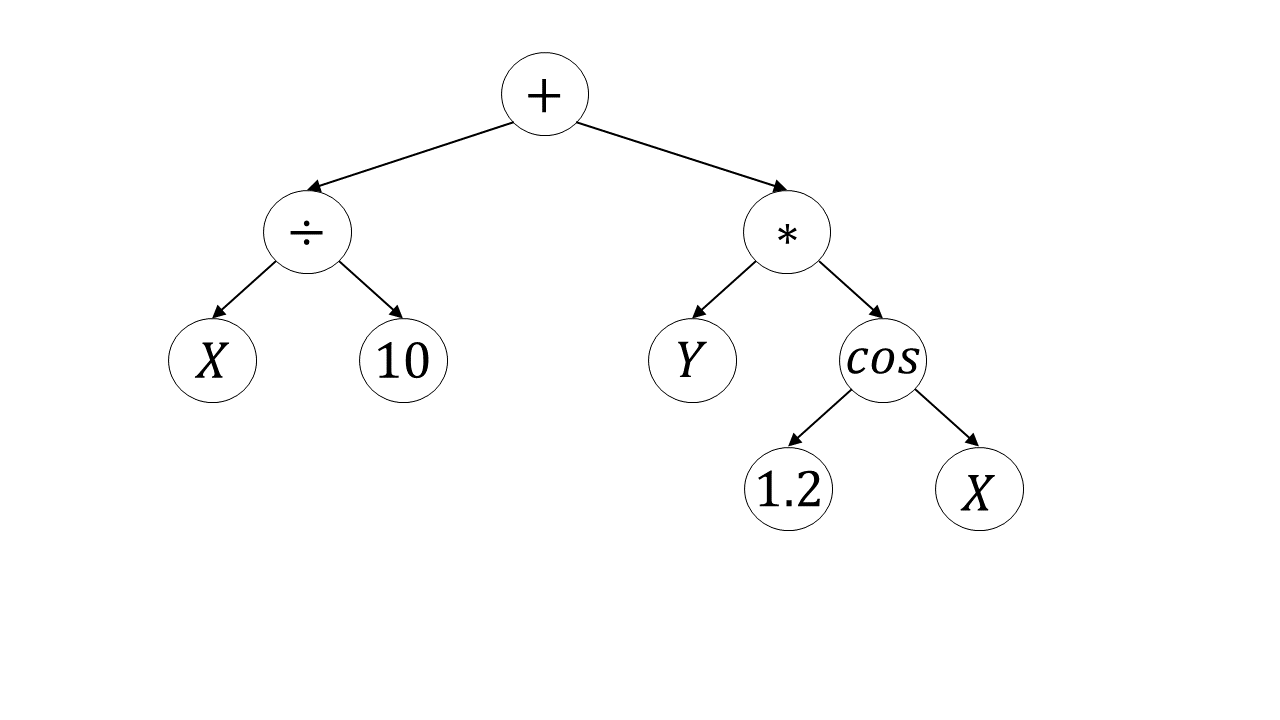}
    \caption{Symbolic representation of $f(X,Y) = \frac{X}{10} + Y * 1.2 \cos (X)$.}
    \label{fig:symbolic}
\end{figure}

Genetic programming relies on a predefined \textit{function set} of mathematical expressions. 
For symbolic regression, the function set typically consists of basic mathematical expressions such as addition, multiplication, and trigonometric terms \citep[see][for details on function set choice]{Nicolau2018}.
Possible model solutions are constructed using a combination of functions from the function set and encoded in a tree structure (Figure \ref{fig:symbolic}).
Within the tree, the mathematical expressions are the decision nodes and input data passed into the mathematical expression are the terminal nodes.
To make the searchable space smaller, the maximum node size of the tree can be specified. 
A \textit{population} of potential solutions is composed of \textit{individual} potential solutions.
The ability of an individual to properly represent data is determined based on the \textit{fitness function}, which is analogous to an objective or loss function in statistics.
Individuals can then \textit{reproduce} to create a copy of themselves, \textit{crossover} with another individual, or \textit{mutate} themselves.
Crossover is where two individuals swap sub-trees (i.e., a decision node is randomly selected from each tree and exchanged) to produce two new individuals, which is equivalent to parents producing offspring with shared genetics.
Mutation is where an individuals decision node is randomly changed (e.g., plus to multiplication or plus to a variable), which is akin to a genetic mutation.

The general algorithm for genetic programming proposes an initial population, assesses the fitness of each individual, and then generates the next population based on the fittest individuals of the current population (Algorithm \ref{alg:genetic}).
Taking the fitness function to be based on regression where the goal is to minimize mean squared error \citep{Schmidt2009}, results in symbolic regression.
This basic genetic programming/symbolic regression method has generated multiple extensions \citep{Icke2013, Chen2017, AmirHaeri2017, Jin2019} and incurred extensive discussion \citep{Korns2014, Nicolau2018, Ahvanooey2019}.

Within the context of data-driven discovery, symbolic regression attempts to find the evolution function $M(\cdot)$ in (\ref{eqn:pde}) or (\ref{eqn:ode}). 
The difficulty is relating a proposed choice for $M(\cdot)$ that is generated within the genetic algorithm to derivatives of the observed data.
Specifically, because derivatives of the system are unknown, either the fitness function needs to account for the derivative or the derivatives must be obtained in order to use a traditional fitness function.

\citet{Bongard2007} were the first to apply symbolic regression to data-driven discovery of dynamic systems, focusing on the discovery of ODEs.
In order to use symbolic regression to discover dynamic models with potentially nonlinear interactions of multiple variables, the authors introduced partitioning, automated probing, and snipping within a symbolic regression algorithm.
Partitioning considers each variable in a system separately, even though they may be coupled, substantially reducing the search space of possible equations.
With partitioning, a candidate equation for a single variable is integrated with the others assumed fixed.
Automated probing is where initial conditions used for temporal integration of the dynamic equation of the system are found.
Last, snipping is the process of simplifying and restructuring models by replacing sub-expressions (sub-trees) in the generated population with a constant.
Using these three components, each variable in the system is integrated forward in time to produce a ``test'' based on the initial condition and compared to the observed data.
The fitness of each potential solution is computed based on the average absolute difference between the observed data and the test.
The approach is illustrated on simulated data and on two real-world examples - the classic hare-lynx system and data they collect from a pendulum.
Although effective, their method is sensitive to noise in the data and has the same demanding computational requirements as other symbolic regression algorithms.

\citet{Schmidt2009} adopt a different approach to data-driven discovery with symbolic regression.
They search over a function space constrained by a loss function dependent on partial derivatives computed from the symbolic functions and from the data.
Specifically, given two variables observed over time, $x(t)$ and $y(t)$ (i.e., $\vu(t) = [x(t), y(t)]'$), the numerical estimate of the partial derivatives between the pair is approximated as $\frac{\Delta x}{\Delta y} \approx \frac{dx}{dt}/\frac{dy}{dt}$, where $\frac{dx}{dt}$ and $\frac{dy}{dt}$ are estimated using local polynomial fits \citep{Thompson1998}. 
From a potential solution function (i.e., generated in the genetic algorithm), the partial derivatives can be computed using symbolic differentiation to get $\frac{\delta x}{\delta y}$ (i.e., from the symbolic function).
To determine how well the potential function expresses the data, the mean log error between the approximated and symbolic partial derivatives,
\begin{align*}
    -\frac{1}{N}\sum_{i=1}^N \log \left( 1 + abs\left(\frac{\Delta x}{\Delta y} - \frac{\delta x}{\delta y} \right) \right),
\end{align*}
is used as the fitness function.
While not discussed here, the approach can be extended to systems with more than two variables by looking at pairs of the variables in the system \citep[see supplementary material of][for details]{Schmidt2009}.
In this manner, they assign a fitness to each proposed individual based on how well the derivative of the system relates to the derivative of the data, resulting in data-driven discovery using symbolic regression.
However, noise can be impactful because the derivatives from the observed data are approximated numerically.
To accommodate measurement uncertainty, \citet{Schmidt2009} use Loess smoothing \citep{Cleveland1988} prior to fitting to remove the high frequency noise.
Their approach is illustrated using simulated data and data collected by motion tracked cameras, showing an ability to recover the equations on complex, real-world problems.
However, similar to \citet{Bongard2007}, the method is also computationally cumbersome with some examples reportedly taking days to converge.

Motivated by symbolic and sparse regression, \citet{Maslyaev2019} embed sparse regression within the coefficient estimation step in a symbolic regression algorithm to discover the governing equations of PDEs.
In their approach, derivatives of the data are computed \textit{a priori} using finite difference (in the same manner as sparse regression discussed in Section \ref{sec:sparse}) and used as the response in symbolic regression.
Within the symbolic regression algorithm, after a population has been proposed, sparse regression using an $\ell_1$ penalty is employed, the fitness of each individual in the population is assessed, and mutation, crossover, and replication are performed in the usual manner.
Because derivatives are computed before the estimation procedure, they are able to be incorporated into the function set.
This allows for the discovered equations to contain spatial derivatives.
The approach is tested on multiple simulated PDEs with varying amounts of measurement noise.
However, the robustness to measurement noise is dependent on the numerical method used to approximate the derivative, and it is unclear how this impacts model results.
Additionally, while specifics are not given, the approach is computationally cumbersome, owing in part to the symbolic regression.

\section{Deep Models}\label{sec:deep}

Deep modeling has been considered for data-driven dynamic discovery in two different ways -- approximating dynamics and learning dynamics.
Approximating dynamics using deep models provides a computationally cheap method to generate data from complex systems while still preserving physical aspects of the system (i.e., emulation).
While this review is concerned with the discovery of the governing equations and refers to ``data-driven discovery'' as the discovery of the \textit{functional form} of the governing system, deep models approximating the dynamics are an important part of the literature and we devote a section to them.
Deep models coinciding with our definition of data-driven discovery have also been developed.
There are multiple approaches by which dynamics can be approximated and subsequently learned, which we also discuss. 

\subsection{Approximating Dynamics with Deep Models}\label{sec:deep_approximate}

One method of approximating dynamics considers a so-called \textit{physics-informed neural network} \citep[PINN;][]{Raissi2017, Raissi2017a, Raissi2018d, Raissi2019, Raissi2020}.
PINNs are applicable to both continuous and discrete time models, yet we discuss only the continuous version here.
Define
\begin{align*}
    \vg(\vs, t) = \vu_{t^{(J)}}(\vs, t) + M\left(\vu(\vs, t), \vu_x(\vs, t), ... \right),
\end{align*}
where the form of $M$ is assumed known.
Approximating $\vu(\vs, t)$ with a neural network results in the PINN $\vg(\vs,t)$, where the derivatives associated with the PINN are computed using automatic differentiation.
The neural network is trained using the loss function $MSE = MSE_u + MSE_g$ where $MSE_u$ is the mean squared error of the neural network approximating $\vu(\vs, t)$ and $MSE_g = \frac{1}{N_g}\sum_{i=1}^{N_g}\|\vg(\vs_i, t_i)\|^2$ is the mean squared error associated with the structure imposed by $\vg(\cdot)$.
In this manner, the neural network obeys the physical constraints imposed by $\vg(\cdot)$.

Neural networks have also been used to approximate the evolution operator $M$ using a residual network (ResNet).
Reframing the problem according to the Euler approximation $\vU(t+\Delta t) \approx \vU(t) + \Delta t M(\vU(t))$, the goal is to find a suitable approximation for $M()$, there-by approximating the dynamics.
In contrast to PINN, physics are not incorporated into the NN and the structure of the NN is dependent completely on the data.
Applying the problem to ODEs, \citet{Qin2019} show how a recurrent ResNet with uniform time steps (i.e., uniform $\Delta t$) and a recursive ResNet with adaptive time steps can be used to approximate dynamics.
This approach is further extended to PDEs \citep{Wu2020}, where the evolution operator is first approximated by basis functions and coefficients, and a ResNet is fit to the basis coefficients.

While not described in detail here, there are other approaches to approximating DE using deep models.
Physics-informed candidate functions can be used with numerical integration in an objective function to restrict the temporal evolution of a NN \citep{Sun2019}.
NN have also been used to approximate parametric PDEs \citep{Khoo2021}, represent molecular dynamics \citep{Mardt2018}, learn turbulent dynamical systems \citep{Qi2022}, and approximate ODEs with time-varying measurement data \citep{Wu2019}.
Using deep models to learn the time-stepping scheme of DE, \citet{Rudy2019a} and \citet{Liu2022} independently show how the approximation of the dynamics are improved when dynamic time-stepping is accounted for in the estimation proceedure.
\citet{Wikle2022} review methods using deep learning for spatial processes and for approximating spatio-temporal DEs.

\subsection{Discovering Dynamics with Deep Models}\label{sec:deep_discover}

Deep modeling using neural networks (NNs) has become increasingly popular in recent years due to NNs being a universal approximator \citep{Hornik1989}.
Additionally, computing derivatives of NNs is possible through automatic differentiation \citep[e.g., using PyTorch;][]{Paszke2017}.
Assuming a surface can be approximated using a NN, derivatives of the surface in space or time or both are obtainable.
This approach, where derivatives are computed using NN, is used in many of the deep model approaches to data-driven discovery.

\subsubsection{Deep Models with Sparse Regression}\label{sec:deep_sparse}

A common issue with data-driven discovery in the ``classical'' sparse regression approach is the sensitivity to noise when approximating derivatives numerically.
To address this issue, \citet{Both2021} propose using a NN to approximate the system, and then perform sparse regression within the NN.
Let $\widehat{\vU}$ be the output of a NN and construct $\vF$ in (\ref{eqn:ode_linear_system}) using $\widehat{\vU}$ and derivatives computed from $\widehat{\vU}$ via automatic differentiation.
The NN is trained using the loss function
\begin{align*}
    \mathcal{L} = \frac{1}{ST}\sum |\vU - \widehat{\vU}|^2 + \frac{1}{ST}\sum |\vF \vM - \widehat{\vU}_{t^{(J)}}|^2 + \lambda \sum |\vM|.
\end{align*}
After training the NN and estimating parameters, most terms of $\vM$ are still nonzero (but very close to zero), and a thresholding is performed to obtain the final sparse representation.
Through this formulation of the problem whereby derivatives are obtained from a NN, \citet{Both2021} show their ability to recover highly corrupt signals from traditional PDE systems and apply their approach to a real-world electrophoresis experiment.

\subsubsection{Deep Models with Symbolic Regression}\label{sec:deep_symbolic}

Using symbolic regression with a neural network for data-driven discovery has gained popularity in recent years. 
In a series of papers, \citet{Xu2019, Xu2020, Xu2021} construct a deep-learning genetic algorithm for the discovery of parametric PDEs (DLGA-PDE) with sparse and noisy data.
DLGA-PDE first trains a NN that is used to compute derivatives and generate meta-data (global and local data), thereby producing a complete de-noised reconstruction of the surface (i.e., noisy sparse data are handled through the NN).
Using the \textit{local} metadata produced by the NN, a genetic algorithm learns the general form of the PDE and identifies which parameters vary spatially or temporally.
At this step, the coefficients may be incorrect or missrepresent the system because the global structure of the data is not accounted for.
To correct the coefficient estimates, a second NN is trained using the discovered structure of the PDE and the \textit{global} metadata.
Last, a genetic algorithm is used to discover the general form of the varying coefficients.

One method of implementing symbolic regression within a deep model is to allow the activation functions to be composed of the function set instead of classic activation functions \citep[e.g., sigmoid or ReLU;][]{Martius2016, Sahoo2018, Kim2021}.
Motivated by this idea, \citet{Long2019} propose a symbolic regression NN, \textit{SymNet}.
Similar to a typical NN, the $\ell$th layer of \textit{SymNet} is
\begin{align*}
    \vf^{\ell} = \vW^{\ell} [\vf^{0}, \vf^{\ell-1}] + \vb^{\ell},
\end{align*}
where $\vf^{0}$ is the function set that contains partial derivatives (e.g., $\vf^{0} = [u, u_x, u_y, ...]$). 
In this manner, each subsequent layer adds a dimension to the activation function based on the previous layer, allowing the construction of complex functions.
Following \citet{Long2017}, spatial derivatives are computed using finite-difference via convolution operators.
To model the time dependence of PDEs, they employ the forward Euler approximation, termed a $\delta t$-block, as
\begin{align*}
    \vU(t + \delta t) \approx \vU(t) + \delta t \cdot SymNet_m^k(u, u_x, u_y, ...),
\end{align*}
where $\delta t$ is the temporal discritization, and $SymNet_m^k(u, u_x, u_y, ...)$ has $k$ hidden layers (i.e., $\ell = 0, ..., k$) and $m$ variables (i.e., number of arguments $u, u_x, u_y, ...$).
In order to facilitate long-term predictions, they train multiple $\delta t$-blocks as a group so the system has long-term accuracy.

Distinct from the previous two approaches, \citet{Atkinson2019} incorporate differential operators into the function set of a genetic algorithm.
They train the NN on the observed data and supply the NN to a genetic algorithm where the function set contains typical operators (e.g., addition, multiplication) and differential operators.
The differential operators are computed from the NN using PyTorch \citep{Paszke2017}, enabling the inclusion of derivatives in the search space of the genetic algorithm.

\section{Physical Statistical Models}\label{sec:psm}

To account for observational uncertainty and missing data when modeling complex non-linear systems, dynamic equations (DE) parameterized by ordinary and partial differential equations have been incorporated into Bayesian hierarchical models (BHM).
While there are various methods by which to model DE in a probabilistic framework, here we focus on physical statistical models \citep[PSM;][]{Berliner1996, Royle1999, Wikle2001} due to the similarities with data-driven discovery that will soon become apparent.
Broadly, PSM are a class of BHMs where scientific knowledge about some process is known and incorporated into the model structure.

PSMs are generally composed of three modeling stages -- data, process, and parameter models -- where dynamics are modeled in the process model and the observed data are modeled conditioned on the latent dynamics.
That is, the observed data are considered to be a noisy \textit{realization} of the ``true'' latent dynamic process.
This formulation results in the data being described conditionally given the process model, simplifying the dependence structure in the data model and enabling complex structure to be captured in the process stage.
The evolution of the latent dynamic process is then parameterized by a DE, incorporating physical dynamics into the modeling framework.

Consider the $R(t) \times 1$ observed data vectors $\vV(t) \equiv [v(\vr_1, t), ..., v(\vr_{R(t)}, t)]'$ and $\{v(\vr, t): \vr \in D_s, t \in D_t\}$ where $\vr \in \{\vr_1, ..., \vr_{R(t)}\} \subset D_s$ is a discrete location in the spatial domain $D_s$, and $t \in \{1, ..., T\} \subset D_t$ is the realization of the system at discrete times in some temporal window $D_t$.
Assume we are interested in the latent ``true'' dynamic process $\{u(\vs, t): \vs \in D_s, t \in D_t\}$ where $\vU(t) \equiv [u(\vs_1, t), ..., u(\vs_S, t)]'$ is a length $S$ vector.
It is common that the observation locations do not coincide with the process (e.g., due to missing data or mismatch in resolution).
In the case of missing observations, the observed data are mapped to the latent process using an incidence matrix $\vH(t)$, which is a matrix of zeros except for a single one in each row corresponding the the observation associated with a process location \citep[see Chapter~7 of][for examples of $\vH(t)$]{Cres}.
The general data model for time $t$ is
\begin{align*}
    \vV(t) = \vH(t)\vU(t) + \veta(t),
\end{align*}
where $\vH(t) \in \mathbb{R}^{L(t) \times N}$ and uncertainty in the observations of the process are captured by $\veta(t) \overset{\text{indep.}}{\sim} N_{L(t)}(\vec{0}, \vSigma_V(t))$ where $\vSigma_V(t)$ is the covariance matrix.

The dynamic process is characterized through the specification of the evolution of $\vU(t)$ over time.
For example, the process model, which specifies this evolution under a first-order Markov assumption, is given as
\begin{align}\label{eqn:gen_process}
    \vU(t) = M(\vU(t-1), \vtheta) + \vepsilon(t),
\end{align}
where $M(\cdot)$ is some (non)linear function relating a previous space-time location (or multiple locations) to the next, $\vtheta$ are parameters associated with $M$, and $\vepsilon(t) \overset{\text{i.i.d.}}{\sim} N(\vec{0}, \vSigma_U)$ is a mean zero Gaussian process with variance/covariance matrix $\vSigma_U$.
While not discussed here, the error term $\vepsilon(t)$ can be considered multiplicative \citep[see Chapter~7 of][for more detail]{Cres}.

Physical dynamics are encoded through the parameterization of $M$.
Here, we consider physical dynamic parameterizations (i.e., ODEs and PDEs), but a general autoregressive structure for $M$ (i.e., not parameterized with differential equations) can also be considered.
Consider the general PDE
\begin{align*}
    \vU_t(t) = M(\vU(t), \vtheta),
\end{align*}
analogous to the motivating PDE (\ref{eqn:pde}), which can be approximated using finite differences
\begin{align*}
    \vU(t) = \vU(t - 1) + \Delta t M(\vU(t - 1), \vtheta),
\end{align*}
where $\Delta t$ is the difference in time between time $t$ and $t-1$ and $\vtheta$ are parameters associated with the PDE.
Because the finite difference approximation can be written as a linear system, we can write
\begin{align}\label{eqn:finite_d}
    \vU(t) = \vM \vU(t - 1),
\end{align}
where $\vM$ is a sparse matrix derived from the finite difference scheme.
Replacing (\ref{eqn:gen_process}) with (\ref{eqn:finite_d}), the process model parameterized by a linear finite difference equation is
\begin{align*}
    \vU(t) = \vM \vU(t - 1) + \vepsilon(t),
\end{align*}
where $\vepsilon(t)$ may now account for approximation error due to the finite difference approximation.

As a clarifying example, assume a spatio-temporal process $U(x,t)$ in one-dimensional space $0 \leq x \leq L$ and time $t$.
Assume the process is approximated by the diffusion equation $U_t(x,t) = b U_{xx}(x,t)$ where $b$ is a diffusion constant and the boundary conditions $Y(0,t) = U_0$ and $U(L,t) = U_L$ and initial condition $\{U(0,t): 0 \leq x \leq L\}$ are known.
Using numerical analysis, the time derivative can be approximated using the forward difference
\begin{align*}
    U_t(x,t) \approx \frac{U(x, t + \Delta t) - U(x, t)}{\Delta t},
\end{align*}
and the spatial derivative can be approximated by the central difference
\begin{align*}
    U_{xx}(x,t) \approx \frac{U(x + \Delta x, t) - 2U(x, t) + U(x - \Delta x, t)}{\Delta x^2}.
\end{align*}
Using the finite difference approximation, we can reformulate the diffusion equation as
\begin{align*}
    U(x, t + \Delta t) \approx U(x, t) + \frac{b \Delta t}{\Delta x^2} \left( U(x + \Delta x, t) - 2U(x, t) + U(x - \Delta x, t) \right).
\end{align*}
Assuming three internal spatial locations, $x_1, x_2, x_3$ and boundary locations $x_0, x_L$, let $\vU(t) = [U(x_1,t), U(x_2,t), U(x_3,t)]'$ and $\vU^b(t) = [U(x_0,t), U(x_L,t)]'$.
Then,
\begin{align*}
    \vU(t + \Delta t) \approx 
    \begin{bmatrix}
        1 - \frac{2 b \Delta t}{\Delta x^2} & \frac{b \Delta t}{\Delta x^2} & 0 \\
        \frac{b \Delta t}{\Delta x^2} & 1 - \frac{2 b \Delta t}{\Delta x^2} & \frac{b \Delta t}{\Delta x^2} \\
        0 & \frac{b \Delta t}{\Delta x^2} & 1 - \frac{2 b \Delta t}{\Delta x^2}
    \end{bmatrix}
    \vU(t) +
    \begin{bmatrix}
        \frac{b \Delta t}{\Delta x^2} & 0 \\
        0 & 0 \\
        0 & \frac{b \Delta t}{\Delta x^2}
    \end{bmatrix}
    \vU^b(t),
\end{align*}
which can be written more compactly as $\vU(t + \Delta t) \approx \vM \vU(t) + \vM^b \vU^b(t)$.  Thus, the PDE dynamics have been ``encoded'' into the structure of the transition operator, $\vM$.
In most PSM implementations, the (banded) structure of $\vM$ is retained, but the specific elements are estimated from the data, rather than given by the finite difference representation.
This adds flexibility and explicitly assumes that the PDE is not an exact representation of the data.
Note that other PDE representations, such as finite element, or spectral, can be used to motivate such models.

This simple example can be made more complex by considering a parametric diffusion equation (i.e., resulting in $\vM(\vtheta)$ instead of $\vM$) or by placing priors on the boundary conditions and or the initial condition \citep[see][for details]{Cres}.
Additionally, there are certain numerical conditions that need to be satisfied in order to guarantee numerical stability from the approximation, which can vary based on the system and approximation scheme considered \citep[e.g., see CFL condition in][]{Higham2016a}.
For a more complete overview of PSMs and possible parameterizations, see \citet{Berliner2003}, \citet{Cres}, \citet{Kuhnert2017}, and references within.

PSMs have been used to study a variety of real-world systems.
PSMs parameterized using shallow-water equations \citep{Wikle2003a} and the Rayleigh friction equation \citep{Milliff2011} have been used to study ocean surface winds.
Using a parametric diffusion equation \citep{Wikle2003a} and parametric reaction-diffusion equation \citep{Hooten2008}, PSMs have modeled the spread of invasive avian species.
PSMs can be grouped into a larger category of models called general quadratic nonlinear model \citep[GQN;][]{Wikle2010, Wikle2011, Gladish2014}, which accommodate multiple classes of scientific-based parameterization such as PDEs and integro-difference equations.

\subsection{General Quadratic Nonlinear Models}\label{sec:GQN}

General quadratic nonlinear models provide a nice generalization to the PSM framework and, as discussed in the subsequent section, provide an interesting link between data-driven discovery methods and PSMs.
The general GQN model is
\begin{align}\label{eqn:gqn_base}
    u(\vs_i, t) = \sum_{j=1}^{S}a_{ij}u(\vs_j, t-1) + \sum_{k=1}^{S}\sum_{l=1}^{S}b_{i, kl}u(\vs_k, t-1)g(u(\vs_l, t-1); \vtheta) + \epsilon(\vs_i, t),
\end{align}
for $i = 1, ..., S$, where $a_{ij}$ are linear evolution parameters, $b_{i, kl}$ are nonlinear evolution parameters, $g()$ is some transformation function of $u(t-1)$ dependent on parameters $\vtheta$, and $\epsilon(\vs_i, t)$ is an error process.
The motivation here is that many real-world mechanistic processes have been described by PDEs that have quadratic (nonlinear) interactions, often where the interaction of system components consists of the multiplication of one component times a transformation of another \citep[see][for details]{Wikle2010}.

Equation (\ref{eqn:gqn_base}) can be condensed in matrix form as
\begin{align}\label{eqn:gqn}
    \vU(t) = \vA\vU(t-1) + (\vI_S \otimes g(\vU(t-1);\vtheta)')\vB\vU(t-1) + \vepsilon(t),
\end{align}
where $\vA$ and $\vB$ are matrices constructed from $a_{ij}$ and $b_{i,kl}$, respectively, and $\vI_S$ is a size $S$ identity matrix \citep[see][for specific details]{Wikle2010}.
From (\ref{eqn:gqn}), we see that letting
\begin{align*}
    \vM(\vU(t-1), \vtheta) = \vA\vU(t-1) + (\vI_S \otimes g(\vV(t-1);\vtheta)')\vB\vU(t-1)
\end{align*}
recovers the PSM model.
The GQN framework is very flexible, due in part to the over-parameterization of the model from all possible quadratic interactions.
To constrain the parameter space, thereby learning which dynamic interactions are important, either physics-informed priors or strong shrinkage priors are used.
For examples on what these constraints may be and their underlying physical motivation, see \citet{Wikle2010}.

\subsection{Relation to Data-Driven Discovery}\label{sec:psm_discovery}

While unexplored in the literature, there is a strong connection between PSMs (particularly, the more general GQNs) and data-driven discovery.
Formulating a BHM where the latent process evolves according to the generic PDE (\ref{eqn:pde}), the two-stage data-process model for location $\vs$ and time $t$ is
\begin{align*}
    \vv(\vs, t) & = \vH(\vs, t)\vu(\vs, t) + \vepsilon(\vs, t) \\
    \vu_{t^{(J)}}(\vs, t) & = M(\vu(\vs, t), \vu_x(\vs, t), ...) + \vepsilon(\vs, t),
\end{align*}
where $\vepsilon(\vs, t) \sim N(\vec{0}, \vSigma_V(\vs, t))$ is the measurement error process with $\vSigma_V(\vs, t)$ a variance/covariance matrix, $\vepsilon(\vs, t) \sim N(\vec{0}, \vSigma_U(\vs, t))$ the process model error process with $\vSigma_U(\vs, t)$ a variance/covariance matrix.
However, as discussed in Section \ref{sec:psm}, PSMs rely on $M$ to be parameterized according to known dynamics.
Instead, borrowing the notion of a feature library from the sparse regression approach to data-driven discovery, linearizing the process model results in a matrix of coefficients $\vM$ and a feature library $\vf(\cdot)$.
Given the feature library, the goal is to find the correct values of $\vM$ (as in sparse regression).
In the case of GQN, we rarely need the whole set of quadratic interactions, so the ``discovery'' is selecting which quadratic components are needed to describe the data.

As an example, consider two approaches that can be used to incorporate dynamic discovery into PSMs - employing a finite difference scheme or using (\ref{eqn:sindy_system}) for the process model -- each of which have their own pros and cons.
The finite difference approach results in the same model as in Section \ref{sec:psm},
\begin{align}\label{eqn:psm_fd}
\begin{split}
    \vv(\vs, t) & = \vH(\vs, t)\vu(\vs, t) + \vepsilon(\vs, t) \\
    \vu(\vs, t) & = \vM \vf(\vu(\vs, t-1), \vu_x(\vs, t-1), ...) + \vepsilon(\vs, t),
\end{split}
\end{align}
where $\vM$ represents the coefficients associated with the finite difference and the discovered equation.
Directly incorporating (\ref{eqn:sindy_system}) in the process model results in
\begin{align}\label{eqn:psm_full}
\begin{split}
    \vv(\vs, t) & = \vH(\vs, t)\vu(\vs, t) + \vepsilon(\vs, t) \\
    \vu_{t^{(J)}}(\vs, t) & = \vM \vf(\vu(\vs, t), \vu_x(\vs, t), ...) + \vepsilon(\vs, t),
\end{split}
\end{align}
where now the temporal derivative is directly related to a library of potential functions and $\vM$ represents the coefficients associated only with the discovered equation.

The benefit of formulating the problem using (\ref{eqn:psm_fd}) is that a Kalman filter or ensemble Kalman filter can be used to estimate parameters
(see \citealt{Stroud2018}, \citealt{Katzfuss2020}, and \citealt{Pulido2018} for examples of the Kalman filter with dynamic systems in statistics).
Additionally, as mentioned previously, the GQN framework naturally provides a construction of an over-parameterized library of potential dynamical interactions into the library.
However, interpreting parameters can be difficult and incorporating spatial derivatives into the library is not as straightforward as with traditional PSMs.
In contrast, (\ref{eqn:psm_full}) has a very clear interpretation of parameters but requires a method to relate the previous state to the current state (e.g., numerical differentiation scheme).
Additionally, model estimation will rely on Metropolis-Hastings Monte-Carlo as the Markov assumption required for Kalman filter and EnKF methods is violated.
For both approaches, parameter shrinkage or variable selection or both will need to be employed on $\vM$ to produce a sparse solution set.
The field of Bayesian variable selection is quite large and there are a variety of priors that can be used \citep[see][for possible choices]{George1993, Park2008, Carvalho2010, Li2010}

Assuming model estimation is possible, formulating the problem using either (\ref{eqn:psm_fd}) or (\ref{eqn:psm_full}) provides significant contributions to the data-driven discovery.
In contrast to the sparse regression approaches with uncertainty quantification discussed in Section \ref{sec:bayes_uncertainty}, (\ref{eqn:psm_fd}) and (\ref{eqn:psm_full}) treat the latent process $\vu(\vs, t)$ as a random process and do not disregard the measurement noise when estimating the system.
That is, instead of computing derivatives and de-noising prior to model estimation, uncertainty in the derivatives as a product of measurement noise is accounted for.
This makes estimation more challenging as the derivatives are no longer assumed known \textit{a priori}.
Additionally, missing data can be handled through the incidence matrix $\vH$.
By formulating the problem within a BHM, known methods accounting for missing data can be used, providing more real-world applicability than the deterministic counterparts.

\section{Bayesian Dynamic Discovery}\label{sec:pdetection}

In a sequence of papers, \citet{North2022, North2022a} propose a fully probabilistic Bayesian hierarchical approach to data-driven discovery of dynamic equations.
Similar to the sparse regression approached discussed in Section \ref{sec:sparse}, the Bayesian approach uses a library of potential functions to identify the governing dynamics.
However, in contrast to the methods discussed in Sections \ref{sec:sparse}, \ref{sec:symbolic}, and \ref{sec:deep}, the dynamic system is modeled as a random process and assumed latent. 
Specifically, \citet{North2022, North2022a} use the approach detailed by (\ref{eqn:psm_full}), where the process model in the BHM,
\begin{align}\label{eqn:sto_system}
    \vu_{t^{(J)}}(\vs, t) & = \vM \vf(\vu(\vs, t), \vu_x(\vs, t), ...) + \vepsilon(\vs, t),
\end{align}
directly relates the time derivative of the dynamic system to a library of potential functions.

In its most general form, (\ref{eqn:sto_system}) has three dimensions, space ($S$), time ($T$), and the number of components ($N$), and can be represented as a tensor -- a higher-order representation of a matrix \citep[see][for a details]{Kolda2006}.
Let $\EuScript{U} = \{u(\vs, t, n): \vs \in D_s, t = 1, ..., T, n = 1, ..., N\}$ where $\EuScript{U} \in \mathbb{R}^{S \times T \times N}$ is the tensor of the dynamic process.
Similarly, let $\EuScript{F}\in \mathbb{R}^{S \times T \times D}$ be the tensor of the function $\vf(\cdot)$ evaluated at each location in space-time and $\widetilde{\veta} \in \mathbb{R}^{S \times T \times N}$ the space-time-component uncertainty tensor.
The process model can represented compactly using tensor notation as
\begin{align}\label{eqn:process_tensor}
    \EuScript{U}_{t^{(J)}} = \EuScript{F} \times_3 \vM + \widetilde{\veta}.
\end{align}
While not explicitly stated, $\EuScript{F}$ is still a function of the state process $\EuScript{U}$ and its derivatives.

The process $\EuScript{U}$ can be defined using a finite collection of spatial, temporal, and component basis functions.
Let
\begin{align*}
    \EuScript{U} & \approx \sum_{p=1}^P \sum_{q=1}^Q \sum_{r=1}^R a(p,q,r) \vpsi(p) \circ \vphi(q) \circ \vtheta(r) = \EuScript{A} \times_1 \vPsi \times_2 \vPhi \times_3 \vTheta \coloneqq [\![ \EuScript{A}; \vPsi, \vPhi, \vTheta ]\!],
\end{align*}
where $\EuScript{A} \in \mathbb{R}^{P \times Q \times R}$, $\vPsi \in \mathbb{R}^{S \times P}$, $\vPhi \in \mathbb{R}^{T \times Q}$, and $\vTheta \in \mathbb{R}^{N \times R}$.
Here, $\vPsi, \vPhi$, and $\vTheta$ are matrices of spatial, temporal, and component basis functions, respectively, and $\EuScript{A}$ is a tensor of basis coefficients (traditionally called the \textit{core tensor}).
Defining $\vPsi$ and $\vPhi$ to be matrices of basis functions differentiable up to at least the highest order considered in (\ref{eqn:pde}), derivatives of $\EuScript{U}$ can be obtained analytically by computing derivatives of the basis functions.
For example
\begin{align*}
    \frac{\partial^3}{\partial x \partial y \partial t} \EuScript{U} = \EuScript{U}_{xyt} = \EuScript{A} \times_1 \vPsi_{xy} \times_2 \vPhi_t \times_3 \vTheta = [\![ \EuScript{A}; \vPsi_{xy}, \vPhi_t, \vTheta ]\!].
\end{align*}
Defining the compact tensor representation of the dynamic process (\ref{eqn:process_tensor}) using the basis decomposition, we can write
\begin{align}\label{eqn:process_basis}
    [\![ \EuScript{A}; \vPsi, \vPhi_{t^{(J)}}, \vTheta ]\!] = \EuScript{F} \times_3 \vM + \veta,
\end{align}
where $\veta$ may now include truncation error.
While not explicitly stated, the arguments of $\EuScript{F}$ are now $\vPsi, \vPhi, \vTheta, \EuScript{A}$, and appropriate derivatives of $\vPsi$ and $\vPhi$.

Taking the mode-3 matricization of (\ref{eqn:process_basis}) -- the flattening of a tensor to a matrix \citep[see][for detail on this operation]{Kolda2006} -- yields
\begin{align*}
    \vTheta \vA (\vphi_{t^{(J)}}(t) \otimes \vpsi(\vs))' = \vM \vf(\vA, \vpsi(\vs), \vpsi_x(\vs), \vpsi_y(\vs), \vpsi_{xy}(\vs), ..., \vphi_{t^{(0)}}(t), ..., \vphi_{t^{(J)}}(t), \vomega(\vs,t)) + \veta(\vs,t),
\end{align*}
where $\vA$ is a $R \times PQ$ matrix of basis coefficients, $\vpsi(\vs)$ is a length-$P$ vector of spatial basis functions, $\vphi(t)$ is a length-$Q$ vector of temporal basis functions, and $\vTheta$ is a $N \times R$ matrix of component basis functions \citep[see][for a detailed explanation]{North2022a}.
This form is convenient because only $\EuScript{A}$ (or $\vA$) needs to be estimated in order to fully define the dynamic process and its derivatives as opposed to requiring all spatial, temporal, or spatio-temporal derivatives of the process, 

Equation (\ref{eqn:pde}) can be extended by rewriting the left-hand side (LHS) to accommodate spatio-temporal derivatives of the process (e.g., $\nabla^2\vu_t(\vs, t) = \vu_{xxt}(\vs, t) + \vu_{yyt}(\vs, t)$), which is common in fluid dynamics \citep[see][for examples]{Higham2016a}.
Specifically, consider the more general PDE
\begin{align}\label{eqn:gen_pde}
    g(\vu_{t^{(J)}}(\vs, t)) = 
    M\left(\vu(\vs, t), \vu_x(\vs, t), \vu_y(\vs, t), \vpsi_{xy}(\vs), ..., \vu_{t^{(1)}}(\vs, t), ..., \vu_{t^{(J-1)}}(\vs, t), \vomega(\vs, t)
    \right),
\end{align}
where $g(\cdot)$ is some linear differential operator.
The basis formulation of (\ref{eqn:gen_pde}) is
\begin{align*}
    \vTheta \vA (\vphi_{t^{(J)}}(t) \otimes g(\vpsi(\vs)))' & = \vM \vf(\vA, \vpsi(\vs), \vpsi_x(\vs), \vpsi_y(\vs), \vpsi_{xy}(\vs), ..., \vphi_{t^{(0)}}(t), ..., \vphi_{t^{(J)}}(t), \vomega(\vs,t)) + \veta(\vs,t),
\end{align*}
where $\veta(\vs, t) \overset{i.i.d.}{\sim} N_N(\vec{0}, \Sigma_U)$ in space and time \citep[see][for a details]{North2022a}.
This results in the general BHM for location $\vs$ and time $t$
\begin{align*}
    \vv(\vs, t) & = \vH(\vs,t) \vTheta \vA (\vphi_{t^{(0)}}(t) \otimes \vpsi(\vs))' + \vepsilon(\vs, t) \\
    \vTheta \vA (\vphi_{t^{(J)}}(t) \otimes g(\vpsi(\vs)))' & = \vM \vf(\vA, \vpsi(\vs), \vpsi_x(\vs), \vpsi_y(\vs), \vpsi_{xy}(\vs), ..., \vphi_{t^{(0)}}(t), ..., \vphi_{t^{(J)}}(t), \vomega(\vs,t)) + \veta(\vs, t),
\end{align*}
where $\vepsilon(\vs, t) \overset{indep.}{\sim} N_{L(\vs,t)}(\vec{0}, \vSigma_V(\vs, t))$ and $\veta(\vs, t) \overset{i.i.d.}{\sim} N_N(\vec{0}, \vSigma_U)$.

Model parameters are estimated by sampling from their full-conditional distributions using Markov chain Monte Carlo, requiring the specification of prior distributions.
Here, we provide a brief summary of the model priors, but for complete model specification, we refer the reader to \citet{North2022} and \citet{North2022a}.
Standard diffuse priors can be assigned to $\vSigma_V(\vs, t)$ and $\vSigma_U$.
In order to induce sparsity in $\vM$, the spike-and-slab prior \citep{Mitchell1988, George1993} is used, where a latent indicator variable $\vgamma = [\gamma_1, ..., \gamma_D]$ denotes if an element of $\vM$ is included in the discovered equation or not.
A constant issue in discovering equations for PDE systems is multicollinearity in the library.
See \citet{North2022a} for a subsampling approach proposed to mitigate the impacts of multicollinearity.
Last, the elastic net prior \citep{Li2010} is assigned to $\vA$ to help regularize the basis coefficients.
Because $\vA$ is embedded in the nonlinear function $\vf(\cdot)$, estimation can be problematic \citep[see][for more detail]{North2022a}.
In order to provide a conjugate updating scheme and reduce computation time, $\vA$ is sampled using an adapted version of stochastic gradient descent (SGD) with a constant learning rate \citep[SGDCL;][]{Mandt2016a}.

While the Bayesian dynamic discovery model proposed in this section relies on more model assumptions and parameters to estimate compared to the methods discussed in Section \ref{sec:sparse}, the benefit is a fully probabilistic discovery of the dynamic system.
The latent variable $\vgamma$ provides an inclusion probability for each element of $\vM$, enabling a researcher to identify a model based on their own desired confidence (i.e., a model identified where each component is included with at least 50\% probability).
Additionally, uncertainty quantification can be obtained for the dynamic system $\vu(\vs, t)$ and all of its subsequent derivatives given the full posterior distribution of $\vA$

\section{Discussion}\label{sec:discussion}

\begin{table}[t]
\begin{threeparttable}
    \centering
    \small
    \begin{tabular}{|l|c|c|c|c|c|c|c|}
        \hline
                                &          & System    &      &     &          & Missing & Real \\
        Reference               & Library  & (ODE/PDE) & Type & UQ  & Noise    & Data    & Data \\
        \hline
        Bongard et al. (2007)   & Symbolic & ODE       & T    & No  & No       & No      & Yes\\
        Schmidt et al. (2009)   & Symbolic & ODE       & T    & No  & Yes      & No      & Yes\\
        Maslyaev et al. (2019)  & Symbolic & PDE       & T    & No  & Yes      & No      & No \\
        Brunton et al. (2016)   & Sparse   & ODE       & T    & No  & Yes      & No      & No \\
        Rudy et al. (2017)      & Sparse   & PDE       & T    & No  & Yes      & No      & No \\
        Rudy et al. (2019)      & Sparse   & PDE       & T    & No  & Yes      & No      & No \\
        Schaeffer (2017)        & Sparse   & PDE       & T    & No  & Yes      & No      & No \\
        Hirsh et al. (2021)     & Sparse   & ODE       & B    & Yes & Yes      & No      & Yes\\
        Zhang et al. (2018)     & Sparse   & PDE       & B    & Yes & Yes      & No\tnote{*} & No \\
        Yang et al. (2020)      & Sparse   & ODE       & B    & Yes & Yes      & No      & No \\
        Bhouri et al. (2022)    & Sparse   & ODE       & B    & Yes & Yes      & No      & No \\
        Fasel et al. (2021)     & Sparse   & PDE       & BO   & Yes & Yes      & No      & Yes\\
        Both et al. (2021)      & Sparse   & PDE       & NN   & No  & Yes      & No      & Yes\\
        Xu et al. (2021)        & Symbolic & PDE       & NN   & No  & Yes      & Yes     & No \\
        Long et al. (2019)      & Symbolic & PDE       & NN   & No  & No       & No      & No \\
        Atkinson et al. (2019)  & Symbolic & PDE       & NN   & No  & No       & No      & Yes\\
        North et al. (2022a)    & Sparse   & ODE       & B    & Yes & Yes      & Yes     & Yes\\
        North et al. (2022b)    & Sparse   & PDE       & B    & Yes & Yes      & Yes     & Yes\\
        \hline
    \end{tabular}
    \begin{tablenotes}\footnotesize
    \item[*] \citet{Zanna2020} applied this framework to real data.
    \end{tablenotes}
    \caption{Summary of some discussed papers where the columns are: \textit{Library} - method used to construct the library, \textit{System} - type of system, either ODE or PDE, considered, \textit{Type} - our categorization of the model (combined with library to get the section it is discussed in) where $T$ is Traditional, $B$ is Bayesian, $BO$ is Bootstrap, and $NN$ is Neural Network, \textit{UQ} - if uncertainty quantification is considered, \textit{Noise} - if the approach considers or can accommodate measurement noise, \textit{Missing Data} - if the approach considers or can accommodate missing data, \textit{Real Data} - if the approach is illustrated using real data.}
    \label{tab:summary}
    \end{threeparttable}
\end{table}

While relatively young, the field of data-driven discovery is expanding quickly.
Areas currently under-studied include methods that properly account for uncertainty quantification and missing data and applications to real-world data sets (see Table \ref{tab:summary}).
BHMs can address these issues, however they rely on the same assumptions as the sparse regression approach - the library is pre-specified.
Relaxing the pre-specified library assumption while retaining the benefits of the statistical approach promises to be a major improvement in the data-driven discovery realm.
To this aim, one promising approach is the recent extension in symbolic regression to the Bayesian framework \citep{Jin2019}.
The incorporation of Bayesian symbolic regression into a BHM could provide the next step to a truly user-free, unbiased, method at data-driven discovery.
Recent advances in deep modeling such as embedding NNs in the BHM \citep{Zammit-Mangion2019} could also be explored.
To this aim, symbolic regression using a NN can be combined with a BHM, providing an alternate method of joining the approaches. 


Real-world data come from a variety of sources such as gridded model output (e.g., reanalysis models), \textit{in situ} observations, and remotely sensed (e.g., satellite) measurements.
While gridded model output is convenient because it is generally complete and spatially and temporally continuous, the discovered dynamics are biased due to the nature of how the data product is constructed.
Conversely, \textit{in situ} and remotely sensed measurements, which are a more direct and unbiased observation of the true dynamic process, are more difficult to use because of they can be missing data and the observations are generally inconsistent in space and time.
Discovery methods that can either use \textit{in situ} measurements, remotely sensed measurements, or both would be beneficial in that the discovered dynamics would be of the ``true'' process and not of the model output.
Additionally, methods combining the model output with \textit{in situ} and remotely sensed measurements, where the spatial and temporal domains may be different (e.g., change of support problem), could provide an extension to not only data-driven discovery but also for change-of-support related methods.

A final direction for future work we discuss are methods that can accommodate mixed data types.
For example, a predator-prey system that takes into account the vegetation coverage of the prey’s spatial domain.
Here, vegetation coverage, a positive continuous variable, is dependent on the number of prey, a positive integer valued variable, which in turn is dependent on the number of predators, another positive integer valued variable.
In a system such as this, the vegetation coverage, which is a positive continuous variable, is dependent on the number of prey, a positive integer valued variable, which in turn is dependent on the number of predators, another positive integer valued variable.
Methods that are able to discover the dynamics of a system with various observed data type provides a much wider range for real-world applications.

\section*{Acknowledgments}
The authors would like to acknowledge Drs.\ Scott Holan and Marianthi Markatou for comments on an early draft.
This research was partially supported by the U.S.~National Science Foundation (NSF) grant SES-1853096 and the U.S. Geological Survey Midwest Climate Adaptation Science Center (CASC) grant No.G20AC00096.

\setstretch{1}
\bibliographystyle{apalike}
\bibliography{references_clean}

\appendix
\section{Algorithms}\label{sec:algorithms}

\SetKwComment{Comment}{/* }{ */} 


\begin{algorithm}
\caption{Sequential Threshold Least-Squares: SINDy}\label{alg:SINDy}
\SetAlgoLined
    \KwIn{$K, \kappa$}
    \KwData{$\vU_{t^{(J)}}, \vF$}
    \KwResult{$\vM$}
    \init{$\vM = (\vF' \vF + \lambda \vI)^{-1} \vF' \vU_{t^{(J)}}$}{}{}
    \For{$k = 1$ \KwTo $K$}{
        $\vgamma = |\vM| < \kappa$ \Comment*[r]{Matrix identifying small coefficients}
        $\vM(\vgamma) = 0$ \Comment*[r]{Threshold $\vM$}
        \For{$n = 1, ..., N$}{
            $i \coloneqq \vgamma(n) == 0$ \Comment*[r]{Identify non-zero columns}
            $\vm(n) = (\vF(i)' \vF(i))^{-1} \vF(i)' \vU_{t^{(J)}}$ \Comment*[r]{Regress}
        }
    }
\end{algorithm}


\begin{algorithm}
\caption{Sequential Threshold Ridge Regression: PDE-FIND}\label{alg:pde_find}
\SetAlgoLined
    \KwIn{$K, \kappa, \lambda$}
    \KwData{$\vU_{t^{(J)}}, \vF$}
    \KwResult{$\vM$}
    \init{$\vM = (\vF' \vF + \lambda \vI)^{-1} \vF' \vU_{t^{(J)}}$}{}{}
    \For{$k = 1$ \KwTo $K$}{
        $\vgamma = |\vM| < \kappa$ \Comment*[r]{Matrix identifying small coefficients}
        $\vM(\vgamma) = 0$ \Comment*[r]{Threshold $\vM$}
        \For{$n = 1, ..., N$}{
            $i \coloneqq \vgamma(n) == 0$ \Comment*[r]{Identify non-zero columns}
            $\vm(n) = (\vF(i)' \vF(i) + \lambda \vI)^{-1} \vF(i)' \vU_{t^{(J)}}$ \Comment*[r]{Regress}
        }
    }
\end{algorithm}


\begin{algorithm}
\caption{Sparse Relaxed Regularized Regression: SR3}\label{alg:sr3}
\SetAlgoLined
    \KwIn{$K, \kappa, \lambda, tolerance$}
    \KwData{$\vU_{t^{(J)}}, \vF, \vW^{0}$}
    \KwResult{$\vM$}
    \init{$k=0$, $err = 2*tolerance$, $\vM = (\vF' \vF + \lambda \vI)^{-1} \vF' \vU_{t^{(J)}}$}{}{}
    \While{$err>tolerance$}{
        $k = k+1$\;
        $\vM^{k} = \underset{\widehat{\vM}}{argmin} \frac{1}{2} \|\vU_{t^{(J)}} - \vF \widehat{\vM}'\|^2 + \frac{1}{2\nu} \|\widehat{\vM} - \vW^{k-1}\|^2$\;
        $\vW^{k} = \text{prox}_{\lambda, \nu, R}(\vM^{k})$ \Comment*[r]{prox is the proximal gradient}
        $err = \|\vW^{k} - \vW^{k-1}\|/\nu$\;
    }
\end{algorithm}


\begin{algorithm}
\caption{General Genetic Algorithm}\label{alg:genetic}
\SetAlgoLined
    \KwIn{Stopping criteria - $\xi$, function set, fitness function - $f()$, summary statistic - $T()$}
    \KwResult{Best individual}
    \init{$P$ = Randomly generate the initial population based on the defined functional set, $\Delta_C = 2\xi, \Delta_N = 0$}{}{}
    \While{$|T(\Delta_C) - T(\Delta_N)| > \xi$}{
        $\Delta_C = f(P)$ \Comment*[r]{Evaluate fitness of current individuals}
        $P$ = Generate new population based on reproduction, crossover, and mutation where individuals are chosen based on fitness level (i.e., higher fitness equals higher probability of being chosen) \;
        $\Delta_N = f(P)$ \Comment*[r]{Evaluate fitness of new individuals}
    }
\end{algorithm}

\end{document}